\documentclass[aps,prl,twocolumn,superscriptaddress,showpacs,preprintnumbers,amsmath,amssymb,floatfix,longbibliography]{revtex4-2}

\usepackage[linktocpage,colorlinks=true,linkcolor=blue,citecolor=blue,breaklinks=true,urlcolor=blue]{hyperref}
\usepackage[usenames,dvipsnames]{xcolor}
\usepackage{graphicx} 
\usepackage{bm} 
\usepackage{color} 
\usepackage{lipsum} 
\usepackage{graphicx}
\usepackage{dcolumn}
\usepackage{bm}
\usepackage{siunitx}
\usepackage{bbold}
\usepackage[mathlines]{lineno}
\usepackage{booktabs}
\usepackage{subfiles}
\usepackage{python}
\usepackage{enumitem}
\usepackage{tikz}
\usepackage{scalerel}
\usepackage[version=4]{mhchem}
\UseRawInputEncoding
\usetikzlibrary{svg.path}

\newcommand{\ecoli}{{\it E.~coli} }

\DeclareSIUnit{\revolutionsperminute}{rpm}

    \definecolor{orcidlogocol}{HTML}{A6CE39}
    \tikzset{
      orcidlogo/.pic={
        \fill[orcidlogocol] svg{M256,128c0,70.7-57.3,128-128,128C57.3,256,0,198.7,0,128C0,57.3,57.3,0,128,0C198.7,0,256,57.3,256,128z};
        \fill[white] svg{M86.3,186.2H70.9V79.1h15.4v48.4V186.2z}
                     svg{M108.9,79.1h41.6c39.6,0,57,28.3,57,53.6c0,27.5-21.5,53.6-56.8,53.6h-41.8V79.1z M124.3,172.4h24.5c34.9,0,42.9-26.5,42.9-39.7c0-21.5-13.7-39.7-43.7-39.7h-23.7V172.4z}
                     svg{M88.7,56.8c0,5.5-4.5,10.1-10.1,10.1c-5.6,0-10.1-4.6-10.1-10.1c0-5.6,4.5-10.1,10.1-10.1C84.2,46.7,88.7,51.3,88.7,56.8z};
      }
    }
    \newcommand\orcid[1]{\href{https://orcid.org/#1}{\mbox{\scalerel*{
    \begin{tikzpicture}[yscale=-1,transform shape]
    \pic{orcidlogo};
    \end{tikzpicture}
    }{|}}}}

\begin{document}

    \title{Selective trapping of bacteria in porous media by cell length}
    
    \author{David Gao}
    \thanks{D.~Gao and R.~Tao contributed equally to this work.}
    \affiliation{Department of Biology,
            University of Pennsylvania, Philadelphia, PA 19104}

    \author{Zeyuan Wang}
    \affiliation{Singh center for nanotechnology,
            University of Pennsylvania, Philadelphia, PA 19104}

    \author{Mihika Jain}
    \affiliation{The Wharton School,
            University of Pennsylvania, Philadelphia, PA 19104}
            
    \author{Arnold J. T. M. Mathijssen\orcid{0000-0002-9577-8928}}
    \email{amaths@upenn.edu}
    \affiliation{Department of Physics \& Astronomy,
            University of Pennsylvania, Philadelphia, PA 19104}

    \author{Ran Tao\orcid{0009-0001-7720-6459}}
    \email{rtao21@upenn.edu}
    \affiliation{Department of Physics \& Astronomy,
            University of Pennsylvania, Philadelphia, PA 19104}
    
    \date{\today}

\begin{abstract}
Bacteria commonly inhabit porous environments such as host tissues, soil, and marine sediments, where complex geometries constrain and redirect their motion. Although bacterial motility has been studied in porous media, the roles of cell length and pore shape in navigating these environments remain poorly understood. Here, we investigate how cell morphology and pore architecture jointly determine bacterial spreading behavior. Using genetically engineered \textit{E.\ coli} with tunable cell length, we performed single-cell tracking in microfluidic devices that mimic ordered and disordered porous structures. We find that elongated bacteria traverse ordered pore networks more effectively than short cells, exhibiting straighter paths, greater directional persistence, and enhanced exploration efficiency. In contrast, in disordered porous media, elongated bacteria become trapped in dead-end regions for extended periods, resulting in markedly reduced navigational efficiency. Together, these results reveal how cell shape and environmental geometry interact to govern bacterial transport. Moreover, we suggest a new mechanism for separating antimicrobial-resistant (AMR) bacteria from elongated susceptible cells in designer porous media. 
\end{abstract}

\textbf{}\maketitle

\section*{INTRODUCTION}

Bacteria commonly inhabit porous environments such as soils and sediments, but also tissues and gut lumen, where complex microstructures constrain and redirect their motion \cite{Philip1970-rf,Adler1988-qy,Battin2016-zh,Baveye2017-bc,Aranson2006-rt,Porter2025-ah,Persat2015-rl,Conrad2012-xr}. These porous landscapes contain networks of tortuous channels punctuated by cracks, cavities, and dead-end regions that strongly influence fluid transport and microbial dispersal \cite{Bhattacharjee2019-ct,Dehkharghani2023-mw,Dentz2022-yt,Creppy2019-ur, Esteves2025-jg, Hallatschek2023-zr, Tao2025-mi}. As a result, the surrounding geometry can critically determine whether bacteria navigate efficiently or become immobilized \cite{Bhattacharjee2019-kg,Dehkharghani2019-lk,Sipos2015-yu,Figueroa-Morales2015-fx,Guasto2012-da}.

Extensive work has characterized bacterial motility in ordered and disordered porous media, in both two and three dimensions, as well as on complex surface topographies \cite{Bhattacharjee2019-ct,Bhattacharjee2019-kg,Creppy2019-ur,De_Anna2021-io, Datta2025-uu,Chang2018-ef,Phan2020-ls,Waisbord2021-lr}. These studies reveal that confinement produces alternating trapping and hopping phases: when a cell encounters a dead end, it undergoes repeated reorientation until an escape path is found, after which it travels persistently along relatively straight channels \cite{Bhattacharjee2019-ct,Dehkharghani2023-mw,Son2013-os}. The relative prevalence of these two phases governs the bacterial transport efficiency and long-time diffusivity, particularly in disordered environments rich in dead ends \cite{Dentz2022-yt,Das2025-qb}.

Cell morphology also plays a central role in modulating motility \cite{Li2008-ge,Jaimes-Lizcano2014-sn,Daddi-Moussa-Ider2020-yk}. Long, filamentous cells typically maintain straighter trajectories and experience more restricted reorientation than short cells, which display classic run-and-tumble behavior with broad turning angles \cite{Guadayol2017-od}. Although both porous geometry and cell shape are known to influence navigation, how these two factors interact within structured environments to regulate transport behavior remains largely unknown \cite{Bhattacharjee2021-oc,Chopra2022-ee}.

Here, we combine inducible cell elongation with precisely engineered microfluidic architectures to investigate how shape--geometry coupling governs bacterial navigation \cite{Perez-Rodriguez2022-bg,Gurung2020-ah}. 
Using genetically modified \textit{E.\ coli} with tunable length \cite{Higashitani1995-qn}, we tracked single-cell trajectories in ordered pillar arrays and disordered porous networks. We find that elongation provides a pronounced advantage in exploring ordered, anisotropic environments. In contrast, within disordered porous media, elongated cells become selectively trapped in cracks and dead-end pockets for extended periods, whereas short cells disperse uniformly. These findings demonstrate that environmental geometry can reverse the benefits of elongation for navigation and reveal a passive mechanism for separating subpopulations, including elongated, antibiotic-stressed cells, based solely on cell length \cite{Galajda2007-ga,Liu2016-jp,Yan2024-mn}.


\section{Results}

To examine how environmental geometry influences cell navigation, we constructed motility chamber and microfluidic devices with three levels of confinement: (i) no confinement, (ii) ordered confinement, and (iii) disordered confinement. The no-confinement condition was created using a motility chamber in which two layers of Parafilm served as a spacer between two coverslips, producing a spacing of \SI{200}{\micro\meter}. Ordered and disordered porous environments were created using microfluidic devices with a depth of \SI{10}{\micro\meter}, enabling continuous imaging of bacteria within the focal plane. The ordered confinement consisted of regularly spaced pillars with a radius $R$ of \SI{25}{\micro\meter} [Methods]. Three versions of the ordered lattice were fabricated, differing only in pillar spacing, which ranged from \SI{10}{\micro\meter} to \SI{50}{\micro\meter}. Moreover, the disordered confinement was composed of irregularly shaped pillars of varying radii fused together to create a heterogeneous pore structure. 

\subsection{Controlling bacterial cell length using inducible gene expression}

\begin{figure*}
    \centering
    \includegraphics[scale = 0.9]{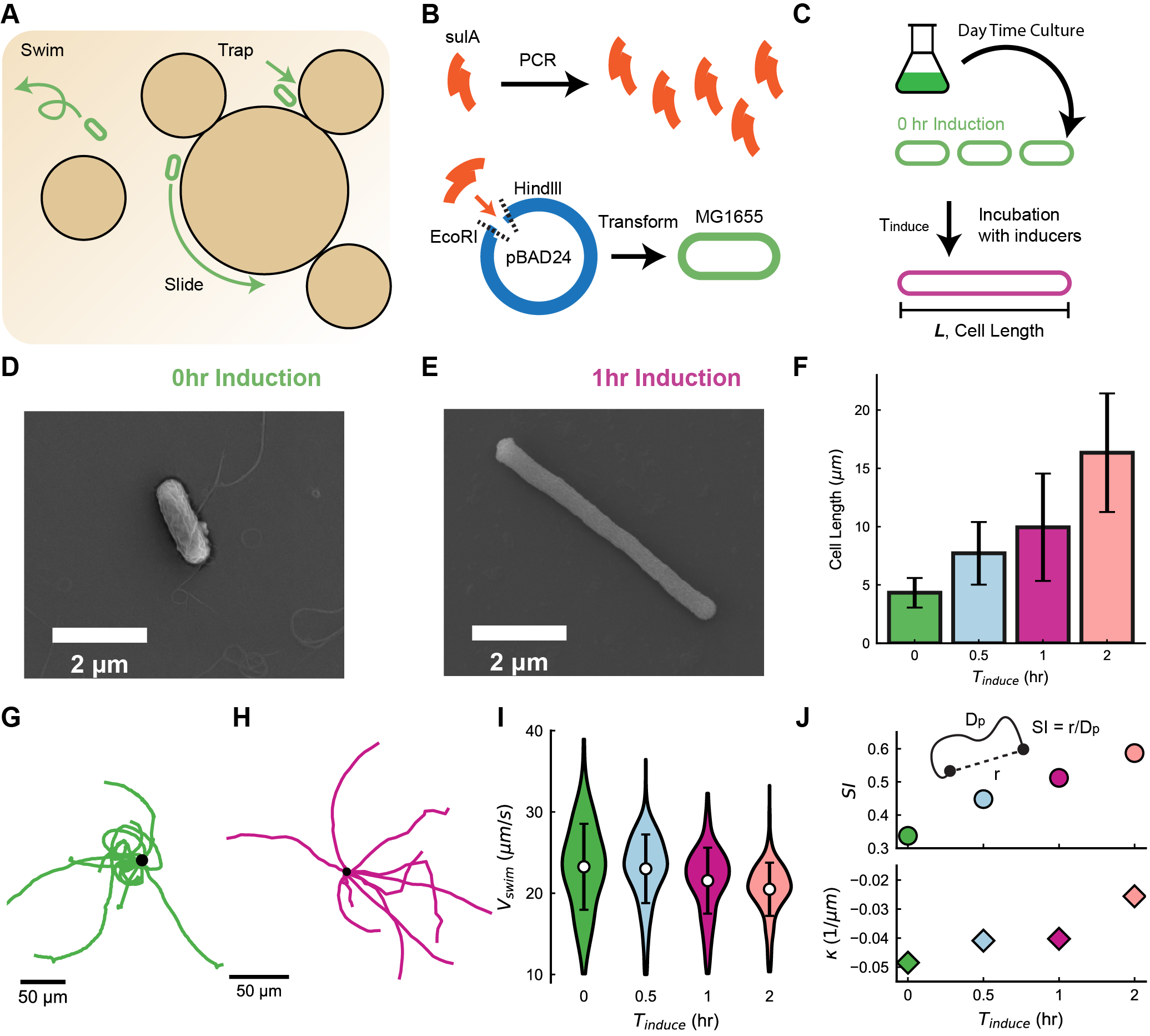}
    \caption{
    \textbf{Bacterial elongation and motility characterization.}
    (A) Schematic illustration of bacterial behaviors in porous media, including free swimming, sliding, and trapping. 
    (B) Genetic construction of the elongation strain. The \textit{sulA} gene is PCR-amplified and inserted into the pBAD24 plasmid using EcoRI and HindIII restriction sites. The resulting recombinant plasmid is transformed into \ecoli MG1655, enabling arabinose-inducible expression of \textit{sulA} for controlled cell elongation.
    (C) Experimental workflow for elongation induction by L-arabinose. 
    (D--E) Representative SEM images of \textit{E.~coli} at 0~hr and 1~hr induction, showing elongation with increasing induction time. 
    (F) Mean cell length measured under bright-field microscopy as a function of induction time $T_{\mathrm{induce}}$. Longer induction results in significantly increased cell length. 
    (G--H) Representative trajectories of short (0~hr, green) and elongated (1~hr, magenta) cells in unconstrained motility chambers. Short cells exhibit circular motion, while elongated cells display straighter, more persistent trajectories. 
    (I) Violin plots of swimming speed $v_{\mathrm{swim}}$ showing minimal dependence on elongation. 
    (J) Straightness index (SI = $r/D_p$), where $r$ is the net displacement and $D_p$ is the total path length, and curvature $\kappa$ as functions of induction time, illustrating that elongated cells swim along straighter trajectories. 
    }
    \label{fig:1}
\end{figure*}

To examine how cell morphology influences motility, we engineered \textit{E.~coli} MG1655 to express the gene \textit{sulA} under the control of the L-arabinose operon (Fig.~\ref{fig:1}B). Induced expression of \textit{sulA} inhibits septum formation, blocking cell division and producing elongated cells (Methods). Following daytime culture, the addition of L-arabinose enabled precise tuning of cell length by varying the induction duration (Fig.~\ref{fig:1}C). SEM images show that non-induced cells remain short rods, whereas induced cells (\SI{1}{\hour} induction time) become markedly elongated (Fig.~\ref{fig:1}D, E). Because dehydration during SEM preparation causes cells to shrink, the quantitative analysis of cell length was performed using bright-field microscopy. These measurements confirm that cell length increases significantly with longer induction times (Fig.~\ref{fig:1}F).

\paragraph{Elongation does not alter swimming speed}

We first examined the motility of elongated bacteria on a flat surface using single-cell tracking. By tracking hundreds of cells ($N>400$), we measured the average swimming speed on a flat surface for each induction time in an unconfined motility chamber near the surface (Methods). The mean swimming speed remained between \SI{21}{\micro\meter\per\second} and \SI{24}{\micro\meter\per\second} in all induction conditions (Fig.~\ref{fig:1}I). Although a slight decrease in measured speed was observed at longer induction times, likely due to reduced tracking accuracy for highly elongated cells, the results indicate that elongation alters cell length but not motility speed, consistent with previous reports \cite{Kamdar2023-eb, Di_Leonardo2011-cz, Kaya2012-po, Maki2000-kp, Zhu2025-as}.

\paragraph{Longer cells explore unconfined surfaces more efficiently}

On flat surfaces, both short and long cells follow curved trajectories, but short cells turn more frequently and exhibit paths with higher curvature. Consequently, short cells tend to remain near their initial positions over extended periods, whereas elongated cells maintain straighter trajectories and explore larger areas (Fig.~\ref{fig:1}G, H). To quantitatively compare exploration efficiency across cell lengths, we defined a dimensionless straightness index (SI), calculated as the ratio between the net displacement (distance between the start and end points of a trajectory) and the total path length traveled. An SI of 0 corresponds to a trajectory that returns to its starting point, representing minimal exploration, while an SI of 1 represents a perfectly straight path and thus the most efficient exploration. Quantitative analysis supports that longer cells exhibit decreased absolute curvature and wider turning radii, and the straightness index increases with induction time (Fig.~\ref{fig:1}J). Together, these results indicate that elongated cells possess greater directional persistence, enabling them to explore unconfined surfaces more efficiently.

\subsection{Navigation of elongated cells in ordered porous media}

\begin{figure*}
    \centering
    \includegraphics[scale = 0.9]{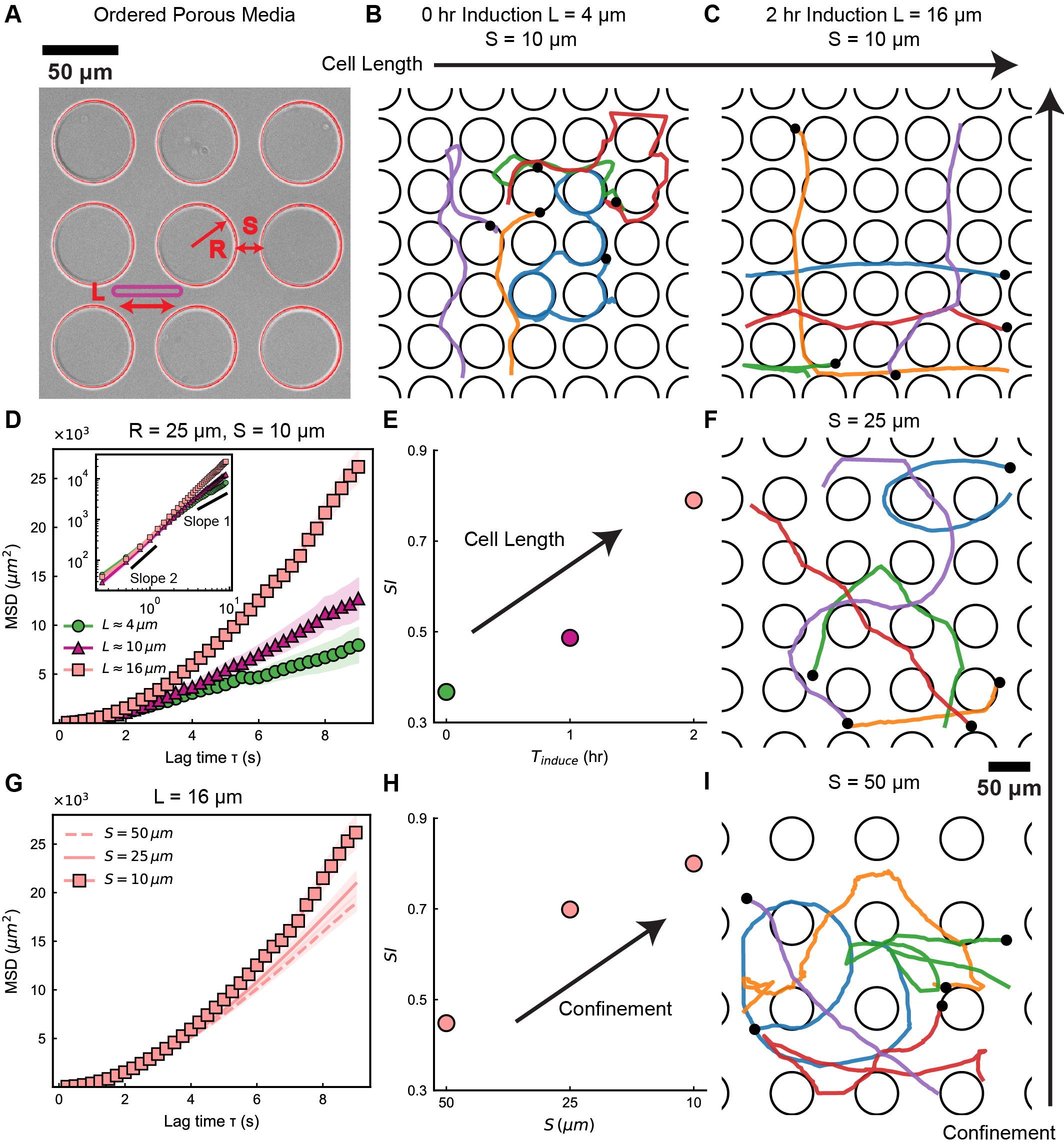}
    \caption{
    \textbf{Effect of cell length and confinement on motility in ordered porous media.}
    (A) Bright-field image of an ordered microfluidic lattice composed of circular pillars with radius $R$ and spacing $S$. 
    (B--C) Trajectories of short ($L = 4~\mu\mathrm{m}$, 0~hr induction) and elongated ($L = 16~\mu\mathrm{m}$, 2~hr induction) cells within ordered pillar arrays ($R = 25~\mu\mathrm{m}$, $S = 10~\mu\mathrm{m}$). Short cells frequently become trapped around pillars, whereas elongated cells display straighter, more persistent trajectories. Each color corresponds to a distinct bacterium.
    (D) Mean-squared displacement (MSD) versus lag time $\tau$ for cells of different lengths under fixed confinement ($R = 25~\mu\mathrm{m}$, $S = 10~\mu\mathrm{m}$); elongated cells exhibit higher MSD values. \textit{Inset:} log--log representation showing reference slopes of $2$ (ballistic) and $1$ (diffusive).
    (E) Straightness index (SI) increases with induction time, indicating enhanced directional persistence with increasing cell length. 
    (C, F, I) Representative trajectories of elongated cells within ordered pillar arrays of varying confinement levels ($S = 10$, 25, and 50~$\mu\mathrm{m}$, respectively). As the pillar spacing increases, trajectories become more curved and less persistent. 
    (G) MSD of elongated cells under different pillar spacings ($S = 10$--$50~\mu\mathrm{m}$), showing enhanced diffusivity under tighter confinement. 
    (H) SI increases with decreasing pillar spacing, consistent with greater trajectory persistence under stronger confinement.
    }
    \label{fig:2}
\end{figure*}

After characterizing the motility of elongated bacteria on an unconfined flat surface, we next examined their behavior in ordered porous media composed of pillars with radius \(R = \SI{25}{\micro\meter}\) and spacing \(S\) ranging from \SI{10}{} to \SI{50}{\micro\meter} (Fig.~\ref{fig:2}A). The microfluidic device was fabricated with a channel height of \(H = \SI{10}{\micro\meter}\), ensuring that cells remained within the focal plane of the microscope throughout the experiments.

\paragraph{Longer cells explore ordered porous media more efficiently} 

Differences in trajectories observed on flat surfaces for bacteria of varying cell lengths persist in highly confined ordered porous media ($R = \SI{25}{\micro\meter}$, $S = \SI{10}{\micro\meter}$). Short cells ($L = \SI{4}{\micro\meter}$) still display highly curved paths and frequently trace around pillar perimeters, leading to inefficient exploration (Fig.~\ref{fig:2}B), agreeing with previous reports \cite{Dehkharghani2023-mw, Chopra2022-ee, Sipos2015-yu}. In contrast, elongated cells move along nearly straight trajectories (Fig.~\ref{fig:2}C). Measurements of the mean squared displacement (MSD) in this same geometry, taken across different cell lengths for hundreds of bacteria ($N>300$), show that longer cells achieve substantially greater spatial expansion over time (Fig.~\ref{fig:2}D). Long cells ($L = \SI{16}{\micro\meter}$) penetrate the ordered porous environment efficiently by passing directly through the gaps between pillars. The pillar geometry restricts their ability to turn, effectively aligning their orientation before sufficient space becomes available for wide turns. Consequently, compared with their shorter counterparts, longer cells in highly confined ordered porous media maintain straight trajectories and achieve high exploration efficiency, as reflected by the SI metric (Fig.~\ref{fig:2}E).

\paragraph{Tighter confinements result in more directed trajectories}

After identifying how cell length influences navigation in ordered porous media, we next examined the effect of pillar spacing \(S\). In highly confined environments, elongated cells exhibit almost perfectly straight trajectories because closely spaced pillars restrict their ability to turn (Fig.~\ref{fig:2}C). As the spacing increases, however, elongated cells encounter more open space, enabling them to execute wide turns and resulting in increasingly curved trajectories (Fig.~\ref{fig:2}I). Mean-square-displacement measurements for long cells across different confinement levels (\(S = \SI{10}{\micro\meter}\), \SI{25}{\micro\meter}, and \SI{50}{\micro\meter}) show that reduced confinement leads to significantly less efficient exploration (Fig.~\ref{fig:2}G). Thus, in addition to cell length, the degree of confinement, set by the spacing between pillars, regulates the exploration efficiency, as captured by the straightness index (Fig.~\ref{fig:2}H). It shows that both the cell length and the confinements levels control bacterial exploration efficiency in porous media.

\subsection{Selective trapping of bacteria based on cell length in disordered porous media}

\begin{figure*}
    \centering
    \includegraphics[scale = 0.9]{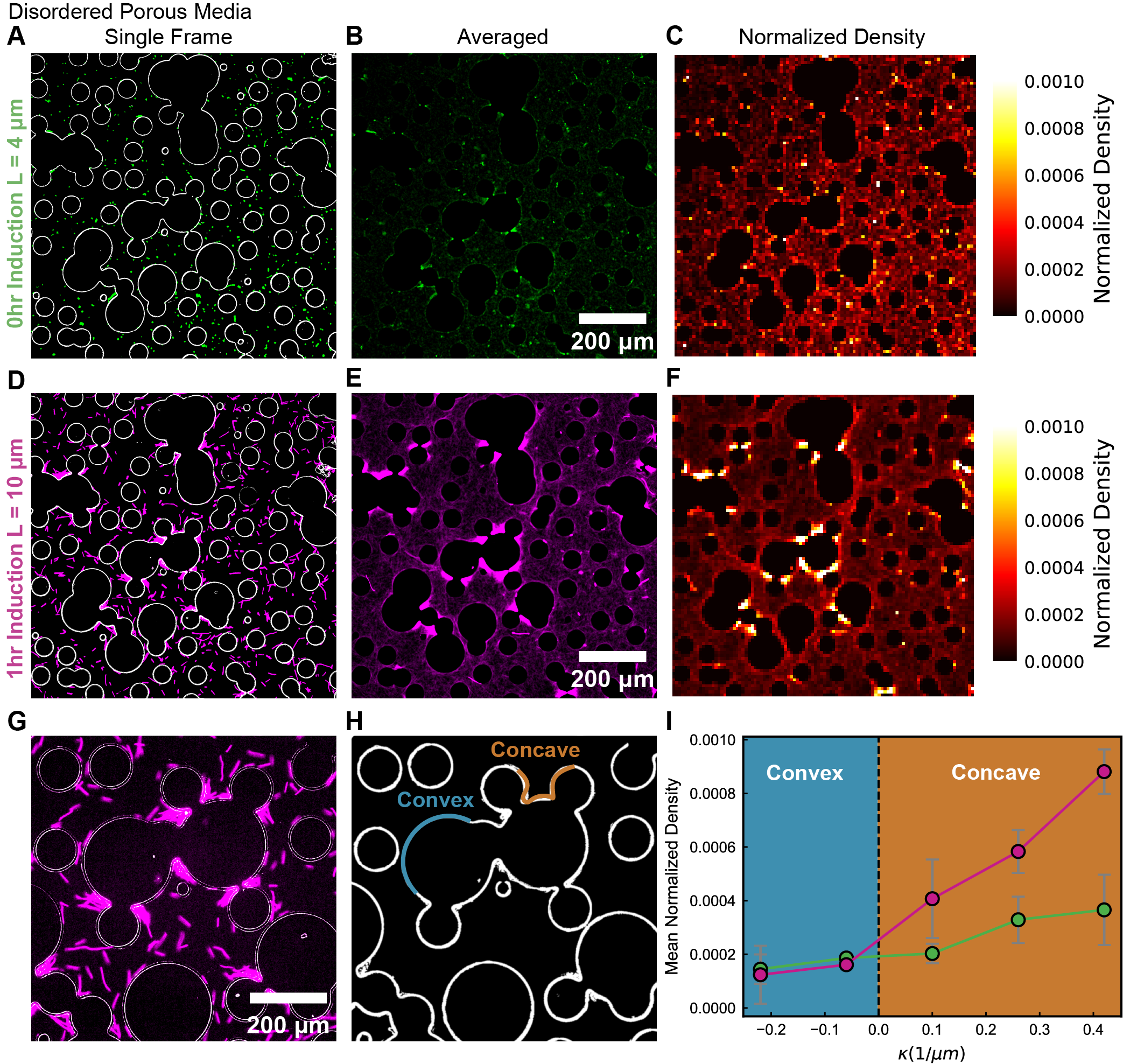}
    \caption{
    \textbf{Trapping of elongated cells in disordered porous media.}
    (A--C) Single-frame, averaged, and normalized density maps for short cells (0~hr induction) in disordered pillar networks, showing a nearly uniform distribution. 
    (D--F) Corresponding maps for elongated cells (1~hr induction), revealing pronounced clustering near dead ends and concave surfaces. 
    (G) Magnified view highlighting elongated cells trapped in concave pore regions. 
    (H) Illustration of concave and convex curvature boundaries used for density analysis. 
    (I) Mean normalized density as a function of local curvature $\kappa$, showing preferential accumulation of elongated cells in concave regions. Positive $\kappa$ corresponds to concave boundaries, and negative $\kappa$ to convex ones.
    }
    \label{fig:3}
\end{figure*}

After characterizing how cells of different lengths navigate ordered environments, we next turned to more natural settings in which obstacles are randomly distributed, forming cracks and dead ends. To test whether the enhanced exploration efficiency of elongated bacteria observed in ordered porous media persists in disordered environments, we inoculated short and long cells, respectively, into a microfluidic device designed to mimic a disordered porous structure (Fig~\ref{fig:3} A). We inoculated short and long cells in the microfluidic device and allowed them to diffuse for \SI{30}{\minute} to ensure sufficient time to respond to the environment. We then acquired time-lapse fluorescence videos of approximately \SI{10}{\minute} to capture the spatial distribution of short and long cells within the disordered porous media. Single frames of time-lapse fluorescence imaging of short and long cells are shown in Fig.~\ref{fig:3}A and Fig.~\ref{fig:3}D, respectively, with white outlines indicating the boundaries of the obstacles. Time-averaged fluorescence images over the full imaging duration are presented in Fig.~\ref{fig:3}B and Fig.~\ref{fig:3}E. The corresponding normalized density distributions are shown in Fig.~\ref{fig:3}C and Fig.~\ref{fig:3}F.

\paragraph{Short cells are uniformly distributed in disordered porous media}

In disordered porous media, short cells were generally uniformly distributed without any clear preference for spatial accumulation (Fig.~\ref{fig:3}A--C). Occasional regions of slightly higher density were attributed to non-motile cells or abnormally long cells present in the culture. Short cells were observed both along obstacle surfaces and within open, unconfined regions. This uniform distribution indicates that short cells can efficiently navigate the disordered environment without becoming trapped at specific locations for extended periods, thereby maximizing their ability to search for nutrients and reducing the likelihood of experiencing local nutrient depletion.

\paragraph{Long cells are trapped in dead ends and cracks}

In contrast to their efficient exploration in ordered porous media, elongated cells preferentially accumulate in dead ends and narrow cracks within disordered porous media (Fig.~\ref{fig:3}D--E). Normalized density maps confirm this localization, showing substantially higher cell density within these confined regions than in the surrounding open areas (Fig.~\ref{fig:3}F). Instead of dispersing uniformly, elongated cells become restricted to these geometric traps, limiting their access to resources. Such confinement is ecologically disadvantageous, as nutrients in dead ends can be rapidly depleted, leading to potential starvation for cells that remain trapped.

\paragraph{Concave microstructures selectively trap elongated cells}

We analyzed the mean normalized density of both short (green) and long (magenta) cells as a function of the local geometric curvature \(\kappa\). For both cell types, higher-curvature regions correspond to increased bacterial accumulation \cite{Perez-Estay2024-ze}, but the effect is substantially more pronounced for elongated cells. This curvature-dependent accumulation explains why short cells remain uniformly distributed throughout the disordered porous medium, whereas elongated cells preferentially accumulate—and become trapped—in concave cracks.

\subsection{Disordered porous media trap elongated cells for prolonged periods}

\paragraph{Elongated cells experience significantly longer trapping times}

\begin{figure*}
    \centering
    \includegraphics[scale=0.9]{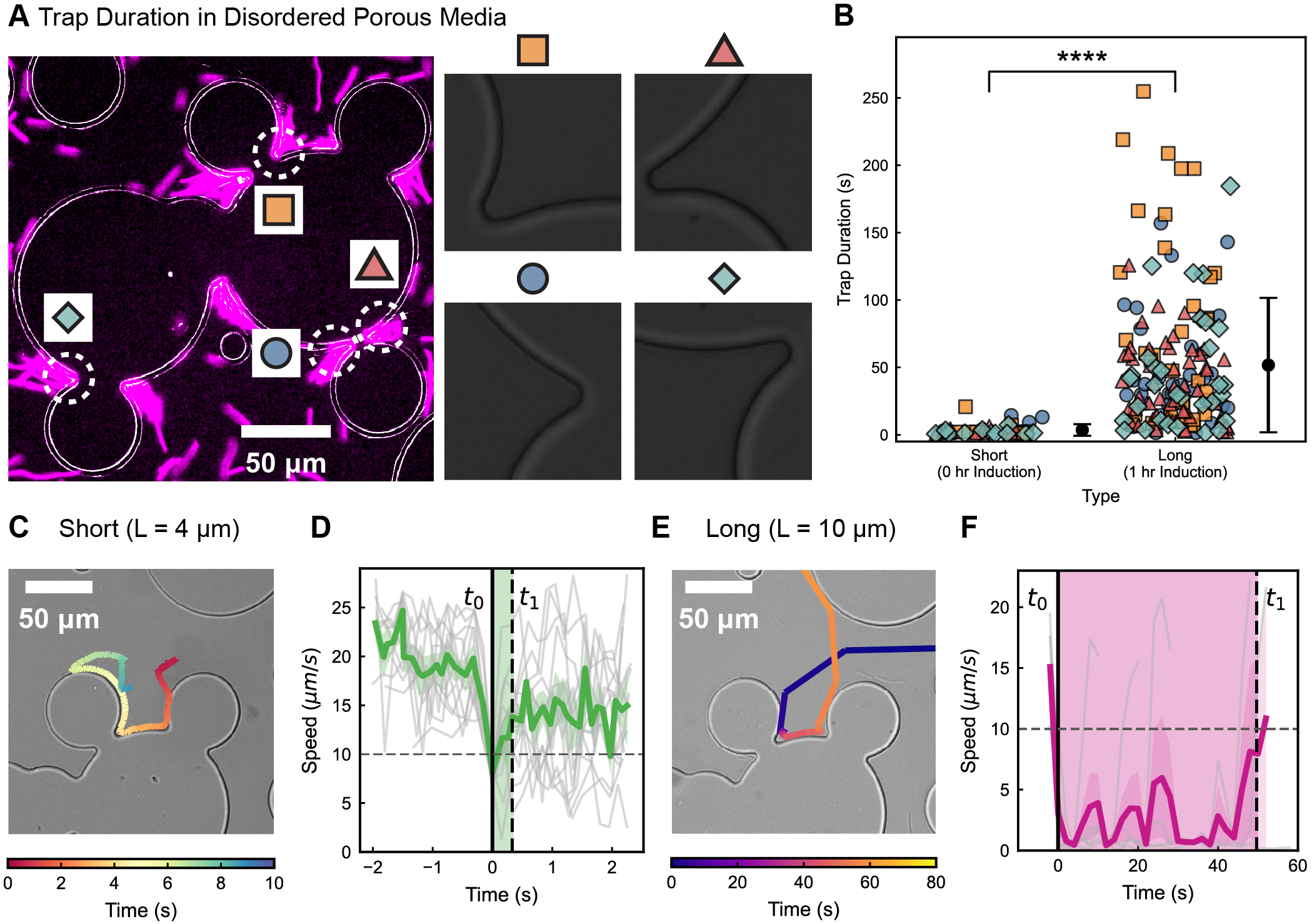}
    \caption{
    \textbf{Trap duration analysis for short and long cells in disordered porous media.}
    (A) Representative regions within disordered porous media used to quantify trap durations, with four example trap locations highlighted.
    (B) Trap duration distributions for short (0 hr induction, $L \approx \SI{4}{\micro\meter}$) and long (1 hr induction, $L \approx \SI{10}{\micro\meter}$) cells. Each point represents an individual trapping event, and different symbols indicate different regions of data collection. A significant difference was observed between groups (****, $p < 0.0001$).
    (C) Example trajectory of a short cell navigating a pillar dead end, color-coded by time.
    (D) Mean speed of short cells aligned to the trap entry time $t_{0}$ and exit time $t_{1}$, with the mean trap duration indicated by shading. Here, $t_{0}$ is defined as the first time the cell speed drops below \SI{10}{\micro\meter\per\second}, marking the start of trapping, and $t_{1}$ is defined as the first time after trapping that the cell speed rises above \SI{10}{\micro\meter\per\second}, marking the end of trapping. 
    (E) Example trajectory of a long cell navigating a pillar dead end. 
    (F) Mean speed of long cells aligned to $t_{0}$ and $t_{1}$, with the mean trap duration indicated by shading.
    }
    \label{fig:4}
\end{figure*}

To understand why elongated cells accumulate densely in concave microstructures, we quantified the trap duration for both short and long cells in representative concave regions. We manually tracked a large number of cells in four distinct concave locations and recorded the trapping time for each trajectory (Fig.~\ref{fig:4}A) ($N = 39$ for short cells and $N = 150$ for long cells). Trap-duration distributions were consistent across all locations, as indicated by the different symbols. Short cells displayed brief trapping events with an average duration of \SI{3.6}{\second}, while elongated cells exhibited a much broader distribution, with mean trapping times of \SI{51.7}{\second}(Fig.~\ref{fig:4}B). These results confirm that elongated cells remain confined within concave regions for periods substantially longer than short cells.

\paragraph{Elongated cells remain motility-limited for extended intervals}

To further probe the mechanistic origin of this prolonged trapping, we examined individual trajectories of short and long cells. Short cells behaved similarly to their motion in ordered porous media. When encountering an obstacle, they quickly reorient themselves to trace around pillars and escape high-curvature trajectories (Fig.~\ref{fig:4}C) \cite{Bhattacharjee2019-ct}. The velocities of individual short-cell trajectories are plotted in gray in Fig.~\ref{fig:4}D. For each trajectory, the first moment at which the velocity fell below \SI{10}{\micro\meter\per\second} was defined as \(t_0\). The dashed line marks the average time at which the velocity subsequently rose above \SI{10}{\micro\meter\per\second}. When averaged across all trajectories, the short cells spent only a brief interval below this velocity threshold before escaping, with an average trapping time \(t_1 - t_0 \approx \SI{0.38}{\second}\), in agreement with the trap-duration distribution shown in Fig.~\ref{fig:4}B.

In contrast, elongated cells were unable to turn sharply enough to trace along the pillar surfaces. When they encountered a dead end, they remained immobilized for extended periods before escaping (Fig.~\ref{fig:4}E--F). Applying the same velocity-based trapping criterion revealed trapping durations of roughly \SI{50}{\second}, again consistent with the distribution measured in Fig.~\ref{fig:4}B. These prolonged intervals of low velocity explain why elongated cells accumulate in concave cracks throughout the disordered porous medium.


\section{Discussion}

\begin{figure*}[t]
    \centering
    \includegraphics[width=\textwidth]{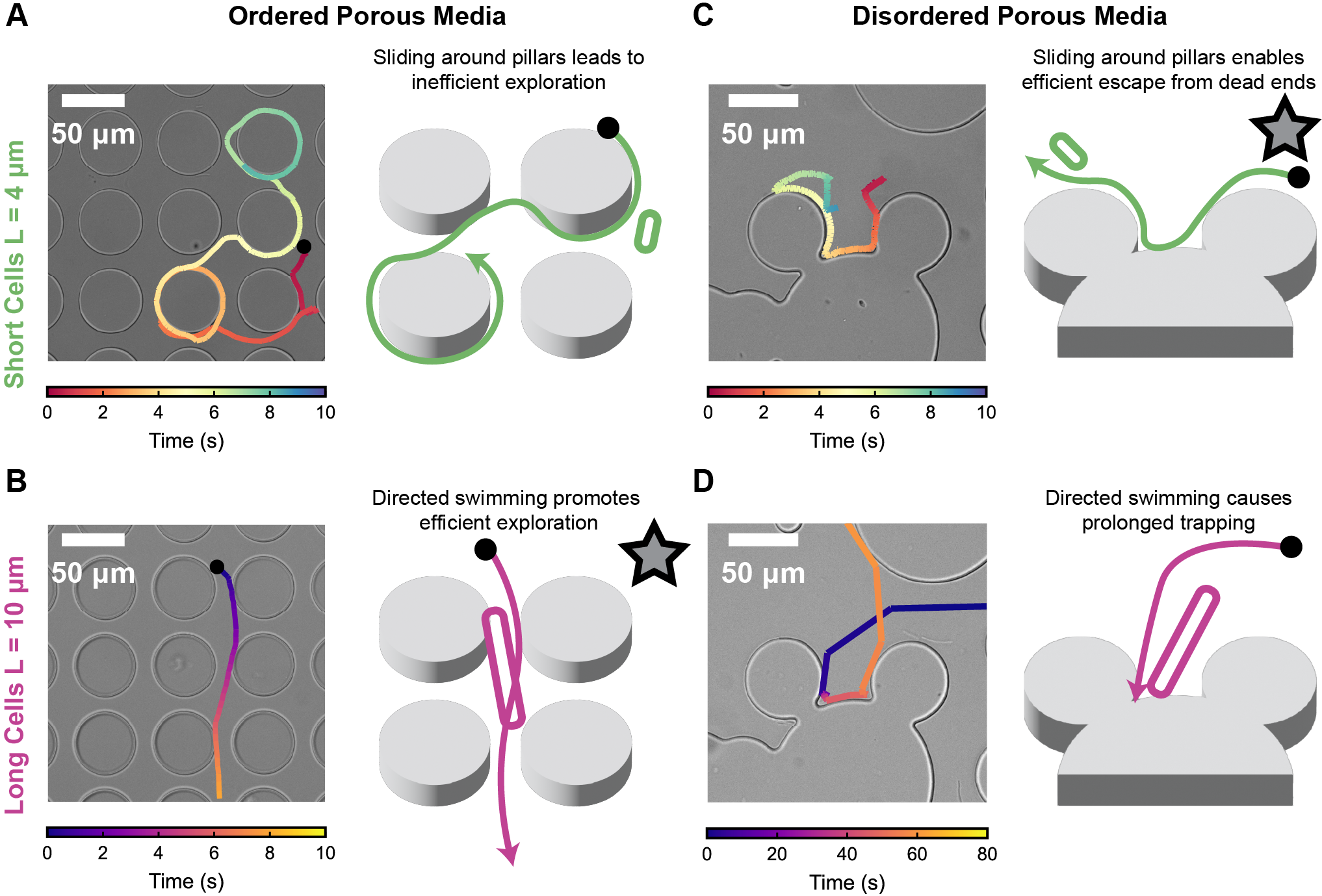}
    \caption{
    \textbf{Navigation of short and long cells in ordered and disordered porous media.}
    (A) In ordered porous media, short cells ($L = \SI{4}{\micro\meter}$) frequently slide along pillar surfaces, resulting in inefficient exploration. 
    (B) Elongated cells ($L = \SI{10}{\micro\meter}$) in ordered media follow directed, low-curvature trajectories, promoting efficient exploration. 
    (C) In disordered porous media, short cells readily escape dead-end regions by sliding along curved pillar boundaries. 
    (D) In contrast, elongated cells become trapped in concave microstructures within disordered porous media, as their limited turning ability prevents effective reorientation. Time–colored trajectories illustrate characteristic navigation patterns, and schematic diagrams highlight the distinct escape and trapping behaviors associated with each condition.
    }
    \label{fig:5}
\end{figure*}

Bacterial navigation within porous environments arises from a complex interplay between cell morphology and microstructural geometry \cite{Philip1970-rf,Bhattacharjee2019-ct,Adler1988-qy}. By combining inducible cell elongation with precisely engineered microfluidic architectures, this study demonstrates that cell length fundamentally reshapes motility strategies across both ordered and disordered porous media. Our central finding is that elongation provides a strong advantage in ordered, anisotropic microstructures but becomes a liability in disordered, heterogeneous environments, where elongated cells are selectively trapped for extended periods. These contrasting behaviors are summarized schematically in Fig.~\ref{fig:5}, which illustrates how short and long cells navigate ordered versus disordered porous structures. Together, these results establish that morphology--geometry coupling governs bacterial transport efficiency and reveal a physical mechanism for sorting or filtering bacteria based solely on cell length \cite{Chopra2022-ee,Figueroa-Morales2015-fx}.

In ordered pillar arrays, elongated cells exhibit markedly enhanced exploratory efficiency due to their persistent, low-curvature trajectories. Because long cells require substantial rotational space to reorient, densely spaced pillars restrict turning and effectively align cell trajectories, allowing elongated bacteria to pass directly through pore corridors with minimal detouring. This behavior is reflected in their higher MSD values (Fig.~\ref{fig:2}D) and larger straightness indices (Fig.~\ref{fig:2}E). In contrast, short cells, whose trajectories are inherently more curved, frequently circle around pillars and reorient, resulting in lower diffusivity \cite{Dehkharghani2019-lk,Sipos2015-yu}. These observations extend prior work on surface motility by showing that environmental anisotropy amplifies the advantages of elongation \cite{Li2008-ge,Galajda2007-ga}.

By keeping cell length constant while varying the degree of geometric confinement, we also demonstrate that exploration efficiency can be tuned purely through environmental structure (Fig.~\ref{fig:2}G--H). This suggests that transport in porous media is controlled by the ratio between a microswimmer's effective length and the local confinement scale. The principles uncovered here are therefore not limited to bacteria. They may apply broadly to other microswimmers and to porous environments across a wide range of length scales \cite{Zottl2019-lt,Zhao2024-og,Tao2021-ra}.

However, the advantages of elongation in ordered lattices reverse completely in disordered porous media, an environment more representative of soils, sediments, and biological tissues. When pore shapes vary irregularly and concave features such as cracks and dead ends are abundant, elongated cells lose the ability to reorient effectively. Unable to turn sharply, they remain immobilized in concave pockets for tens of seconds, nearly one order of magnitude longer than short cells (Fig.~\ref{fig:4}B). Curvature-dependent analysis further reveals that concave microstructures selectively trap elongated cells (Fig.~\ref{fig:3}I), whereas short cells remain uniformly distributed across the network (Fig.~\ref{fig:3}A--C). Thus, the geometric complexity of porous media governs not only whether cells experience trapping, but which subpopulation is trapped. Spatial disorder therefore acts as a natural selective filter against elongated cells \cite{Bhattacharjee2019-kg,Creppy2019-ur,Das2025-qb}.

These results unify and extend previous studies by demonstrating that porous microstructures do not uniformly impede motion. Their influence depends sensitively on cell shape and on the spatial organization of the environment \cite{Dentz2022-yt,De_Anna2021-io,Rusconi2014-cu}. In natural ecosystems, where pore geometry often varies across scales, this morphology-specific behavior likely shapes microbial colonization, nutrient access, and competitive dynamics \cite{Baveye2017-bc,Battin2016-zh}. Filamentation, commonly induced by antibiotic stress in \ecoli and other species, may inadvertently hinder dispersal, trapping elongated cells in unfavorable microhabitats and limiting their ability to access nutrients \cite{Choe2025-fu,Nair2023-hm}. 
Conversely, antimicrobial-resistant (AMR) cells, which often avoid filamentation, may navigate these disordered environments more effectively \cite{Yan2024-mn,Li2024-xo}.

Therefore, a key implication of our findings is that porous systems can be deliberately engineered to separate bacterial subpopulations by length, such as antimicrobial-resistant and susceptible populations. In ordered porous networks, elongated bacteria migrate rapidly through channels while short cells lag behind. In disordered networks, elongated bacteria accumulate in high-curvature concavities while short cells traverse the domain more freely. Such length-selective trapping enables label-free sorting of antibiotic-treated, filamented, or resistant cells. For example, introducing a mixed population into an ordered lattice would cause elongated, antibiotic-stressed cells to reach downstream regions first, enriching resistant cells in specific spatial zones. In contrast, a disordered lattice would trap elongated cells and allow short cells to reach the downstream end more quickly. These passive separation mechanisms could be useful for diagnostics or for studying antibiotic-induced morphological changes \cite{Liu2016-jp,Perez-Rodriguez2022-bg,Gurung2020-ah}.

Beyond single-cell motility, our findings also have implications for early surface colonization and biofilm development in porous environments. Biofilms often initiate in regions of prolonged residence time such as concave boundaries, low-shear pockets, and dead-end pores, where cells experience enhanced surface contact \cite{Donlan2002-ah,Drescher2013-xn,Feng2015-ak}. Flow-induced streamers can rapidly form in these confined regions and clog pore spaces, dramatically altering transport pathways \cite{Drescher2013-xn,Battin2016-zh, Dehkharghani2019-lk,Tao2025-mi}. Moreover, curved or textured surfaces have been shown to bias initial bacterial attachment through hydrodynamic interactions and local flow gradients \cite{Secchi2020-ml,Galajda2007-ga,Rusconi2014-cu, Creppy2019-ur}. Microfluidic studies further demonstrate that geometric confinement, surface topography, and local hydrodynamics play essential roles in shaping both single-cell adhesion and collective biofilm morphology \cite{Perez-Rodriguez2022-bg, Secchi2020-ml}. Taken together, these observations suggest that the geometry-dependent motility behaviors identified in this work may directly influence where biofilms nucleate at the first place and how microbial communities become spatially organized in natural and engineered porous environments.

Several limitations highlight possible studies for future work. First, due to microscopy limitation, our microfluidic devices represent two-dimensional projections of inherently three-dimensional porous habitats. Extending these studies to fully three-dimensional architectures, such as transparent hydrogels or granular media, may reveal additional trapping modes or escape pathways \cite{Kim2010-mt,Kim2022-ky}. Second, chemical gradients, fluid flow, and extracellular polymer production, all common in natural bacterial ecosystems, may further modulate morphology-dependent trapping \cite{Oliveira2022-pc,Feng2015-ak}. Third, collective behaviors such as swarming, alignment interactions, and quorum-sensing-mediated responses could either enhance or suppress the navigation patterns observed here \cite{Donlan2002-ah,Perez-Rodriguez2022-bg}. Lastly, most natural or physiologically relevant porous environments are filled with non-Newtonian fluids \cite{Gonzalez-La-Corte2025-zi, Cao2024-nm, Torres_Maldonado2024-wu}, which can substantially alter bacterial morphology and swimming behavior. It would therefore be interesting to test whether the observations reported in this study persist under non-Newtonian fluid conditions. Incorporating these biological factors into controllable microfluidic systems will deepen our understanding of how morphology shapes microbial movement through complex environments.

Overall, our findings demonstrate that bacterial cell length is not merely a morphological trait but a functional determinant of transport efficiency in heterogeneous media. By showing that environmental geometry can selectively trap elongated cells while facilitating or impeding motion depending on spatial order, this work bridges microbiology, soft matter physics, and microfluidic engineering. The ability to rationally design porous structures that exploit morphology-dependent motility opens new opportunities for microbial sorting, ecological control, and antibiotic-response diagnostics \cite{Gurung2020-ah}.


\small
\section{Methods}

\subsection{Bacterial cultures and genetic modification}

The bacterial strain used in these experiments was \ecoli MG1655 carrying a chromosomal \textit{ompA-cfp} fusion. Tunable cell length in \ecoli was achieved via heterologous expression of \textit{sulA}, encoding a repressor of Z-ring formation that inhibits cell division \cite{Higashitani1995-qn}. The gene \textit{sulA} was PCR amplified from \ecoli MG1655 and cloned into the EcoRI and HindIII sites of pBAD24 following standard restriction enzyme cloning procedures. Use of EcoRI ensured the \textit{sulA} gene was located proximal to the strong ribosome binding site in the pBAD24 MCS. The resulting construct was confirmed via Sanger sequencing, and \ecoli MG1655 was transformed with the plasmid via chemical transformation and plating on LB with \SI{100}{\micro\gram\per\milli\liter} ampicillin. All growth media were supplemented with \SI{0.2}{\percent} D-glucose to repress expression of \textit{sulA}. To induce elongation, glucose was omitted and \SI{0.2}{\percent} L-arabinose was added instead to activate \textit{sulA} expression from the P\textit{BAD} promoter.

Cells from a frozen stock were streaked onto agar plates (1\% Bacto tryptone, 0.5\% yeast extract, 1.0\% NaCl, 1.5\% agar) and incubated overnight at \SI{32}{\degreeCelsius}. Single colonies of \textit{E. coli} from freshly streaked plates were used to inoculate \SI{3}{\milli\liter} of LB medium (1\% Bacto tryptone, 0.5\% yeast extract, 1.0\% NaCl) supplemented with \SI{2}{\micro\gram\per\milli\liter} ampicillin. Cultures were grown overnight at \SI{32}{\degreeCelsius} to stationary phase with shaking at \SI{250}{\revolutionsperminute}. A \SI{100}{\micro\liter} aliquot of the overnight culture was diluted $10^{-2}$ in LB medium (1\% Bacto tryptone, 0.5\% yeast extract, 1.0\% NaCl) supplemented with \SI{2}{\micro\gram\per\milli\liter} ampicillin and incubated with shaking for approximately \SI{3}{\hour}, until the optical density at \SI{600}{\nano\metre} (OD$_{600}$) reached 0.1. The resulting culture was then supplemented with \SI{0.2}{\percent} L-arabinose to induce cell elongation for \SI{0}{}--\SI{2}{\hour}, where induction duration determined the average cell length. Cells with the desired length were subsequently diluted $10^{-2}$ in Motility Buffer (MB: \SI{0.1}{mM} EDTA, \SI{0.001}{mM} L-methionine, \SI{10}{mM} sodium lactate, \SI{67}{mM} NaCl, \SI{6.2}{mM} \ce{K2HPO4}, \SI{3.9}{mM} \ce{KH2PO4}) and supplemented with \SI{0.08}{\gram\per\milli\liter} L-serine and \SI{0.03}{\percent} polyvinylpyrrolidone (PVP). The motility buffer was adjusted to a pH of approximately 7.05. Cells were suspended in the motility buffer for \SI{30}{\minute}, and their swimming behavior was examined in a motility chamber before experiments to ensure optimal adaptation. All microfluidic measurements were conducted within \SI{2}{\hour} at room temperature to maintain stable motility.

All microfluidic experiments were repeated at least three times using freshly cultured bacteria on different days.

\subsection*{Motility chamber and Microfluidic devices}

The motility chamber was constructed from two coverslips separated by a double layer of Parafilm, creating a spacing of \SI{200}{\micro\meter}. The microfluidic channels have a depth of \(H = \SI{10}{\micro\meter}\) and contain pillars of varying radius \(R = \SI{5}{\micro\meter}\)--\SI{50}{\micro\meter} arranged in either ordered or disordered geometries. The microfluidic devices were fabricated from polydimethylsiloxane (PDMS) by replica molding against a positive-relief silicon wafer master coated with SU-8 patterns produced using standard photolithography and soft-lithography procedures. PDMS and curing agent were mixed thoroughly at a 10:1 weight ratio, degassed under vacuum for at least one hour to remove air bubbles, and poured onto the silicon wafer. The mixture was then cured at \SI{65}{\celsius} overnight to ensure full cross-linking. After curing, the PDMS replica was cut, peeled from the master, and inlet and outlet ports of \SI{1}{\milli\meter} diameter were punched for bacterial loading. Both the PDMS replica and a glass coverslip were cleaned with compressed air and plasma-treated for \SI{25}{\second} (Harrick Plasma PDC-32G). The two components were then irreversibly bonded and placed on a \SI{95}{\celsius} hot plate for approximately \SI{1}{\minute} to strengthen the seal.

To design the disordered porous medium, we used the Python-based LayoutScript interface of LayoutEditor to generate an array of circular pores within a rectangular domain of size \SI{1000}{\micro\meter} \(\times\) \SI{4000}{\micro\meter}. The pore array consisted of three radii, \SI{10}{\micro\meter}, \SI{25}{\micro\meter}, and \SI{50}{\micro\meter}, distributed at fractional abundances of \SI{10}{\percent}, \SI{75}{\percent}, and \SI{15}{\percent}, respectively. These pores were initially positioned on a regular grid with center--to--center spacing \(S = \SI{25}{\micro\meter}\). To introduce geometric disorder, we applied independent Gaussian displacements to the \(x\)-- and \(y\)--coordinates of each pore center. Specifically, each pore originally located at \((x_0, y_0)\) was shifted according to a Gaussian distribution with standard deviation

\[
\sigma = \frac{(1 - \gamma) S}{\sqrt{2}},
\]

where \(\gamma\) is the disorder index.

In this formulation, \(\gamma = 1\) corresponds to a perfectly ordered lattice, while smaller \(\gamma\) yields increasingly disordered pore architectures. In this work, we used \(\gamma = 0.1\), which produced strongly heterogeneous porous structures in a controlled and reproducible manner.


\subsection*{Imaging and single-cell tracking}

Bacterial dynamics were imaged on a Nikon TI2-E microscope using a 20X objective (CFI60 Plan Apochromat Lambda, numerical aperture 0.75, working distance \SI{1.0}{\milli\meter}). Bright-field imaging was used for single-cell tracking, and fluorescence imaging was used for cell distribution analysis. Videos were recorded with a scientific CMOS camera (Hamamatsu ORCA-Fusion Gen~III) at frame rates up to 100~FPS.

Microfluidic channels were flushed with bacteria suspended in motility buffer until an optimal cell concentration was reached, dilute enough to prevent cell overlap and interference during tracking, yet concentrated enough to obtain robust datasets. Single-cell tracking was performed to characterize bacterial diffusion in porous media of varying pillar sizes and alignments.

Cells were tracked only while swimming within the focal plane to ensure high-quality trajectories. Videos were processed in ImageJ, and cell positions were extracted automatically using a custom macro built based on the TrackMate plugin. Trajectories lasting longer than \SI{1}{\second} were analyzed in Python to get data used for analysis in this paper.

\normalsize

\section*{Author Contributions}

David Gao and Ran Tao contributed equally to this work. Ran Tao and Arnold J.\ T.\ M.\ Mathijssen conceived the project and designed the experiment. Ran Tao and Zeyuan Wang designed and fabricated the microfluidic devices. David Gao, Ran Tao, and Mihika Jain performed the bacterial experiments. Ran Tao analyzed the data. Ran Tao and David Gao wrote the manuscript draft. Ran Tao and Arnold J.\ T.\ M.\ Mathijssen revised and finalized the manuscript with input from all coauthors. Ran Tao supervised the project. Arnold J.\ T.\ M.\ Mathijssen provided funding.

\section{Acknowledgements}
\begin{acknowledgments}
We are grateful to Nathaniel Esteves and Jay Zhu for providing the bacterial strains and for their valuable help with the inducible cell length protocol. We further thank Liuni Chen and Ling Li for their support with SEM imaging. We thank all members of the Mathijssen lab for their support and insightful discussions. A.J.T.M.M. acknowledges funding from the Charles E. Kaufman Foundation (Early Investigator Research Award, KA2022-129523; New Initiative Research Award, KA2024-144001) and the National Science Foundation (UPenn MRSEC, DMR-2309043).
\end{acknowledgments}


\bibliography{PorousMediaReferencePaperpile}

@ARTICLE{Bhattacharjee2019-kg,
  title    = "Confinement and activity regulate bacterial motion in porous media",
  author   = "Bhattacharjee, Tapomoy and Datta, Sujit S",
  journal  = "Soft Matter",
  volume   =  15,
  number   =  48,
  pages    = "9920--9930",
  month    =  "11~" # dec,
  year     =  2019,
  url      = "http://dx.doi.org/10.1039/c9sm01735f",
  doi      = "10.1039/c9sm01735f",
  pmid     =  31750508,
  issn     = "1744-683X,1744-6848"
}

@ARTICLE{Dehkharghani2019-lk,
  title     = "Bacterial scattering in microfluidic crystal flows reveals giant
               active Taylor-Aris dispersion",
  author    = "Dehkharghani, Amin and Waisbord, Nicolas and Dunkel, Jörn and
               Guasto, Jeffrey S",
  journal   = "Proc. Natl. Acad. Sci. U. S. A.",
  publisher = "Proceedings of the National Academy of Sciences",
  volume    =  116,
  number    =  23,
  pages     = "11119--11124",
  month     =  "4~" # jun,
  year      =  2019,
  url       = "http://dx.doi.org/10.1073/pnas.1819613116",
  doi       = "10.1073/pnas.1819613116",
  pmc       = "PMC6561239",
  pmid      =  31097583,
  issn      = "0027-8424,1091-6490"
}

@ARTICLE{Creppy2019-ur,
  title     = "Effect of motility on the transport of bacteria populations
               through a porous medium",
  author    = "Creppy, Adama and Clément, Eric and Douarche, Carine and
               D'Angelo, Maria Veronica and Auradou, Harold",
  journal   = "Phys. Rev. Fluids",
  publisher = "American Physical Society (APS)",
  volume    =  4,
  number    =  1,
  month     =  "16~" # jan,
  year      =  2019,
  url       = "http://dx.doi.org/10.1103/physrevfluids.4.013102",
  doi       = "10.1103/physrevfluids.4.013102",
  issn      = "2469-990X,2469-9918"
}

@ARTICLE{Li2008-ge,
  title    = "{Amplified effect of Brownian motion in bacterial near-surface
              swimming}",
  author   = "Li, Guanglai and Tam, Lick-Kong and Tang, Jay X",
  journal  = "Proc. Natl. Acad. Sci. U. S. A.",
  volume   =  105,
  number   =  47,
  pages    = "18355--18359",
  month    =  "25~" # nov,
  year     =  2008,
  url      = "http://dx.doi.org/10.1073/pnas.0807305105",
  doi      = "10.1073/pnas.0807305105",
  pmc      = "PMC2587629",
  pmid     =  19015518,
  issn     = "0027-8424,1091-6490"
}

@ARTICLE{Figueroa-Morales2015-fx,
  title     = "{Living on the edge: Transfer and traffic of E. coli in a
               confined flow}",
  author    = "Figueroa-Morales, Nuris and Leonardo Miño, Gastón and Rivera,
               Aramis and Caballero, Rogelio and Clément, Eric and Altshuler,
               Ernesto and Lindner, Anke",
  journal   = "Soft Matter",
  publisher = "Royal Society of Chemistry",
  volume    =  11,
  number    =  31,
  pages     = "6284--6293",
  month     =  "21~" # aug,
  year      =  2015,
  url       = "http://dx.doi.org/10.1039/c5sm00939a",
  doi       = "10.1039/c5sm00939a",
  pmid      =  26161542,
  issn      = "1744-683X,1744-6848"
}

@ARTICLE{Daddi-Moussa-Ider2020-yk,
  title     = "{Tuning the Upstream Swimming of Microrobots by Shape and Cargo
               Size}",
  author    = "Daddi-Moussa-Ider, Abdallah and Lisicki, Maciej and Mathijssen,
               Arnold J T M",
  journal   = "Phys. Rev. Appl.",
  publisher = "American Physical Society",
  volume    =  14,
  number    =  2,
  pages     =  024071,
  month     =  "25~" # aug,
  year      =  2020,
  url       = "https://link.aps.org/doi/10.1103/PhysRevApplied.14.024071",
  doi       = "10.1103/PhysRevApplied.14.024071",
  issn      = "2331-7019,2331-7043"
}

@ARTICLE{Battin2016-zh,
  title    = "The ecology and biogeochemistry of stream biofilms",
  author   = "Battin, Tom J and Besemer, Katharina and Bengtsson, Mia M and
              Romani, Anna M and Packmann, Aaron I",
  journal  = "Nat. Rev. Microbiol.",
  volume   =  14,
  number   =  4,
  pages    = "251--263",
  month    =  apr,
  year     =  2016,
  url      = "http://dx.doi.org/10.1038/nrmicro.2016.15",
  doi      = "10.1038/nrmicro.2016.15",
  pmid     =  26972916,
  issn     = "1740-1526,1740-1534"
}

@ARTICLE{Kamdar2023-eb,
  title    = "Multiflagellarity leads to the size-independent swimming speed of
              peritrichous bacteria",
  author   = "Kamdar, Shashank and Ghosh, Dipanjan and Lee, Wanho and
              Tătulea-Codrean, Maria and Kim, Yongsam and Ghosh, Supriya and
              Kim, Youngjun and Cheepuru, Tejesh and Lauga, Eric and Lim,
              Sookkyung and Cheng, Xiang",
  journal  = "Proc. Natl. Acad. Sci. U. S. A.",
  volume   =  120,
  number   =  48,
  pages    = "e2310952120",
  month    =  "28~" # nov,
  year     =  2023,
  url      = "http://dx.doi.org/10.1073/pnas.2310952120",
  doi      = "10.1073/pnas.2310952120",
  pmc      = "PMC10691209",
  pmid     =  37991946,
  issn     = "0027-8424,1091-6490"
}

@ARTICLE{Phan2020-ls,
  title     = "Bacterial Route Finding and Collective Escape in Mazes and
               Fractals",
  author    = "Phan, Trung V and Morris, Ryan and Black, Matthew E and Do, Tuan
               K and Lin, Ke-Chih and Nagy, Krisztina and Sturm, James C and
               Bos, Julia and Austin, Robert H",
  journal   = "Phys. Rev. X",
  publisher = "American Physical Society",
  volume    =  10,
  number    =  3,
  pages     =  031017,
  month     =  "22~" # jul,
  year      =  2020,
  url       = "https://link.aps.org/doi/10.1103/PhysRevX.10.031017",
  doi       = "10.1103/PhysRevX.10.031017",
  issn      = "2160-3308"
}

@ARTICLE{Kim2022-ky,
  title     = "Effects of swimming environment on bacterial motility",
  author    = "Kim, Dokyum and Kim, Yongsam and Lim, Sookkyung",
  journal   = "Phys. Fluids",
  publisher = "AIP Publishing",
  volume    =  34,
  number    =  3,
  pages     =  031907,
  month     =  mar,
  year      =  2022,
  url       = "https://pubs.aip.org/aip/pof/article/34/3/031907/2844585",
  doi       = "10.1063/5.0082768",
  issn      = "1070-6631,1089-7666"
}

@ARTICLE{Donlan2002-ah,
  title    = "Biofilms: microbial life on surfaces",
  author   = "Donlan, Rodney M",
  journal  = "Emerg. Infect. Dis.",
  volume   =  8,
  number   =  9,
  pages    = "881--890",
  month    =  sep,
  year     =  2002,
  url      = "http://dx.doi.org/10.3201/eid0809.020063",
  doi      = "10.3201/eid0809.020063",
  pmc      = "PMC2732559",
  pmid     =  12194761,
  issn     = "1080-6040"
}

@ARTICLE{Bhattacharjee2019-ct,
  title    = "Bacterial hopping and trapping in porous media",
  author   = "Bhattacharjee, Tapomoy and Datta, Sujit S",
  journal  = "Nat. Commun.",
  volume   =  10,
  number   =  1,
  pages    =  2075,
  month    =  "6~" # may,
  year     =  2019,
  url      = "http://dx.doi.org/10.1038/s41467-019-10115-1",
  doi      = "10.1038/s41467-019-10115-1",
  pmc      = "PMC6502825",
  pmid     =  31061418,
  issn     = "2041-1723"
}

@ARTICLE{Waisbord2021-lr,
  title    = "{Fluidic bacterial diodes rectify magnetotactic cell motility in
              porous environments}",
  author   = "Waisbord, Nicolas and Dehkharghani, Amin and Guasto, Jeffrey S",
  journal  = "Nat. Commun.",
  volume   =  12,
  number   =  1,
  pages    =  5949,
  month    =  "12~" # oct,
  year     =  2021,
  url      = "http://dx.doi.org/10.1038/s41467-021-26235-6",
  doi      = "10.1038/s41467-021-26235-6",
  pmc      = "PMC8511139",
  pmid     =  34642318,
  issn     = "2041-1723"
}

@ARTICLE{Perez-Rodriguez2022-bg,
  title     = "Microfluidic devices for studying bacterial taxis, drug testing
               and biofilm formation",
  author    = "Pérez-Rodríguez, Sandra and García-Aznar, José Manuel and
               Gonzalo-Asensio, Jesús",
  journal   = "Microb. Biotechnol.",
  publisher = "Wiley",
  volume    =  15,
  number    =  2,
  pages     = "395--414",
  month     =  feb,
  year      =  2022,
  url       = "https://sfamjournals.onlinelibrary.wiley.com/doi/abs/10.1111/1751-7915.13775",
  doi       = "10.1111/1751-7915.13775",
  pmc       = "PMC8867988",
  pmid      =  33645897,
  issn      = "1751-7907,1751-7915"
}

@ARTICLE{Rusconi2014-cu,
  title     = "{Bacterial transport suppressed by fluid shear}",
  author    = "Rusconi, Roberto and Guasto, Jeffrey S and Stocker, Roman",
  journal   = "Nat. Phys.",
  publisher = "Nature Publishing Group",
  volume    =  10,
  number    =  3,
  pages     = "212--217",
  month     =  "23~" # feb,
  year      =  2014,
  url       = "https://www.nature.com/articles/nphys2883",
  doi       = "10.1038/nphys2883",
  issn      = "1745-2473,1745-2481"
}

@ARTICLE{Tao2021-ra,
  title    = "Soft particle clogging in two-dimensional hoppers",
  author   = "Tao, Ran and Wilson, Madelyn and Weeks, Eric R",
  journal  = "Phys Rev E",
  volume   =  104,
  number   = "4-1",
  pages    =  044909,
  month    =  oct,
  year     =  2021,
  url      = "http://dx.doi.org/10.1103/PhysRevE.104.044909",
  doi      = "10.1103/PhysRevE.104.044909",
  pmid     =  34781509,
  issn     = "2470-0053,2470-0045"
}

@ARTICLE{Kim2010-mt,
  title    = "Diffusion-based and long-range concentration gradients of multiple
              chemicals for bacterial chemotaxis assays",
  author   = "Kim, Minseok and Kim, Taesung",
  journal  = "Anal. Chem.",
  volume   =  82,
  number   =  22,
  pages    = "9401--9409",
  month    =  "15~" # nov,
  year     =  2010,
  url      = "http://dx.doi.org/10.1021/ac102022q",
  doi      = "10.1021/ac102022q",
  pmid     =  20979359,
  issn     = "0003-2700,1520-6882"
}

@ARTICLE{Zottl2019-lt,
  title     = "Enhanced bacterial swimming speeds in macromolecular polymer
               solutions",
  author    = "Zöttl, Andreas and Yeomans, Julia M",
  journal   = "Nat. Phys.",
  publisher = "Nature Publishing Group",
  volume    =  15,
  number    =  6,
  pages     = "554--558",
  month     =  "11~" # mar,
  year      =  2019,
  url       = "https://www.nature.com/articles/s41567-019-0454-3",
  doi       = "10.1038/s41567-019-0454-3",
  issn      = "1745-2473,1745-2481"
}

@ARTICLE{Feng2015-ak,
  title    = "Bacterial attachment and biofilm formation on surfaces are reduced
              by small-diameter nanoscale pores: how small is small enough?",
  author   = "Feng, Guoping and Cheng, Yifan and Wang, Shu-Yi and Borca-Tasciuc,
              Diana A and Worobo, Randy W and Moraru, Carmen I",
  journal  = "NPJ Biofilms Microbiomes",
  volume   =  1,
  pages    =  15022,
  month    =  "2~" # dec,
  year     =  2015,
  url      = "http://dx.doi.org/10.1038/npjbiofilms.2015.22",
  doi      = "10.1038/npjbiofilms.2015.22",
  pmc      = "PMC5515209",
  pmid     =  28721236,
  issn     = "2055-5008"
}

@ARTICLE{Secchi2020-ml,
  title     = "{The effect of flow on swimming bacteria controls the initial
               colonization of curved surfaces}",
  author    = "Secchi, Eleonora and Vitale, Alessandra and Miño, Gastón L and
               Kantsler, Vasily and Eberl, Leo and Rusconi, Roberto and Stocker,
               Roman",
  journal   = "Nat. Commun.",
  publisher = "Nature Research",
  volume    =  11,
  number    =  1,
  pages     =  2851,
  month     =  "5~" # jun,
  year      =  2020,
  url       = "http://dx.doi.org/10.1038/s41467-020-16620-y",
  doi       = "10.1038/s41467-020-16620-y",
  pmc       = "PMC7275075",
  pmid      =  32503979,
  issn      = "2041-1723"
}

@ARTICLE{Galajda2007-ga,
  title    = "A wall of funnels concentrates swimming bacteria",
  author   = "Galajda, Peter and Keymer, Juan and Chaikin, Paul and Austin,
              Robert",
  journal  = "J. Bacteriol.",
  volume   =  189,
  number   =  23,
  pages    = "8704--8707",
  month    =  dec,
  year     =  2007,
  url      = "http://dx.doi.org/10.1128/JB.01033-07",
  doi      = "10.1128/JB.01033-07",
  pmc      = "PMC2168927",
  pmid     =  17890308,
  issn     = "0021-9193,1098-5530"
}

@ARTICLE{Sipos2015-yu,
  title     = "{Hydrodynamic Trapping of Swimming Bacteria by Convex Walls}",
  author    = "Sipos, O and Nagy, K and Di Leonardo, R and Galajda, P",
  journal   = "Phys. Rev. Lett.",
  publisher = "American Physical Society",
  volume    =  114,
  number    =  25,
  pages     =  258104,
  month     =  "26~" # jun,
  year      =  2015,
  url       = "http://dx.doi.org/10.1103/PhysRevLett.114.258104",
  doi       = "10.1103/PhysRevLett.114.258104",
  pmid      =  26197146,
  issn      = "0031-9007,1079-7114"
}

@ARTICLE{Gurung2020-ah,
  title    = "Microfluidic techniques for separation of bacterial cells via
              taxis",
  author   = "Gurung, Jyoti P and Gel, Murat and Baker, Matthew A B",
  journal  = "Microb. Cell Fact.",
  volume   =  7,
  number   =  3,
  pages    = "66--79",
  month    =  "15~" # jan,
  year     =  2020,
  url      = "http://dx.doi.org/10.15698/mic2020.03.710",
  doi      = "10.15698/mic2020.03.710",
  pmc      = "PMC7052948",
  pmid     =  32161767,
  issn     = "2311-2638"
}

@ARTICLE{Zhao2024-og,
  title     = "Integrating organoids and organ-on-a-chip devices",
  author    = "Zhao, Yimu and Landau, Shira and Okhovatian, Sargol and Liu,
               Chuan and Lu, Rick Xing Ze and Lai, Benjamin Fook Lun and Wu,
               Qinghua and Kieda, Jennifer and Cheung, Krisco and Rajasekar,
               Shravanthi and Jozani, Kimia and Zhang, Boyang and Radisic,
               Milica",
  journal   = "Nat Rev Bioeng",
  publisher = "Springer Science and Business Media LLC",
  volume    =  2,
  number    =  7,
  pages     = "588--608",
  month     =  "2~" # jul,
  year      =  2024,
  url       = "http://dx.doi.org/10.1038/s44222-024-00207-z",
  doi       = "10.1038/s44222-024-00207-z",
  issn      = "2731-6092"
}

@ARTICLE{Liu2016-jp,
  title     = "High-throughput screening of antibiotic-resistant bacteria in
               picodroplets",
  author    = "Liu, X and Painter, R E and Enesa, K and Holmes, D and Whyte, G
               and Garlisi, C G and Monsma, F J and Rehak, M and Craig, F F and
               Smith, C A",
  journal   = "Lab Chip",
  publisher = "The Royal Society of Chemistry",
  volume    =  16,
  number    =  9,
  pages     = "1636--1643",
  month     =  "26~" # apr,
  year      =  2016,
  url       = "https://pubs.rsc.org/en/content/articlelanding/2016/lc/c6lc00180g/unauth",
  doi       = "10.1039/c6lc00180g",
  pmid      =  27033300,
  issn      = "1473-0197,1473-0189"
}

@ARTICLE{Li2024-xo,
  title     = "Dynamics of rigid fibers interacting with triangular obstacles in
               microchannel flows",
  author    = "Li, Zhibo and Bielinski, Clément and Lindner, Anke and du Roure,
               Olivia and Delmotte, Blaise",
  journal   = "Phys. Rev. Fluids",
  publisher = "American Physical Society (APS)",
  volume    =  9,
  number    =  4,
  month     =  "8~" # apr,
  year      =  2024,
  url       = "http://dx.doi.org/10.1103/physrevfluids.9.044302",
  doi       = "10.1103/physrevfluids.9.044302",
  issn      = "2469-990X,2469-9918"
}

@ARTICLE{Choe2025-fu,
  title     = "Rapid identification of key antibiotic resistance genes in {E}.
               coli using high-resolution genome-scale {CRISPRi} screening",
  author    = "Choe, Donghui and Lee, Eunju and Kim, Kangsan and Hwang, Soonkyu
               and Jeong, Ki Jun and Palsson, Bernhard O and Cho, Byung-Kwan and
               Cho, Suhyung",
  journal   = "iScience",
  publisher = "Elsevier BV",
  volume    =  28,
  number    =  5,
  pages     =  112435,
  month     =  "16~" # may,
  year      =  2025,
  url       = "http://dx.doi.org/10.1016/j.isci.2025.112435",
  doi       = "10.1016/j.isci.2025.112435",
  pmc       = "PMC12063145",
  pmid      =  40352728,
  issn      = "2589-0042"
}

@ARTICLE{Nair2023-hm,
  title     = "Interspecies interaction reduces selection for antibiotic
               resistance in Escherichia coli",
  author    = "Nair, Ramith R and Andersson, Dan I",
  journal   = "Commun. Biol.",
  publisher = "Springer Science and Business Media LLC",
  volume    =  6,
  number    =  1,
  pages     =  331,
  month     =  "27~" # mar,
  year      =  2023,
  url       = "https://www.nature.com/articles/s42003-023-04716-2",
  doi       = "10.1038/s42003-023-04716-2",
  pmc       = "PMC10043022",
  pmid      =  36973402,
  issn      = "2399-3642"
}

@ARTICLE{Yan2024-mn,
  title     = "A label-free droplet sorting platform integrating
               dielectrophoretic separation for estimating bacterial
               antimicrobial resistance",
  author    = "Yan, Jia-De and Yang, Chiou-Ying and Han, Arum and Wu, Ching-Chou",
  journal   = "Biosensors (Basel)",
  publisher = "MDPI AG",
  volume    =  14,
  number    =  5,
  pages     =  218,
  month     =  "26~" # apr,
  year      =  2024,
  url       = "http://dx.doi.org/10.3390/bios14050218",
  doi       = "10.3390/bios14050218",
  pmc       = "PMC11117925",
  pmid      =  38785691,
  issn      = "2079-6374"
}

@ARTICLE{Oliveira2022-pc,
  title     = "Suicidal chemotaxis in bacteria",
  author    = "Oliveira, Nuno M and Wheeler, James H R and Deroy, Cyril and
               Booth, Sean C and Walsh, Edmond J and Durham, William M and
               Foster, Kevin R",
  journal   = "Nat. Commun.",
  publisher = "Springer Science and Business Media LLC",
  volume    =  13,
  number    =  1,
  pages     =  7608,
  month     =  "9~" # dec,
  year      =  2022,
  url       = "https://www.nature.com/articles/s41467-022-35311-4",
  doi       = "10.1038/s41467-022-35311-4",
  pmc       = "PMC9734745",
  pmid      =  36494355,
  issn      = "2041-1723"
}

@ARTICLE{Chopra2022-ee,
  title     = "Geometric effects induce anomalous size-dependent active
               transport in structured environments",
  author    = "Chopra, Pooja and Quint, David and Gopinathan, Ajay and Liu, Bin",
  journal   = "Phys. Rev. Fluids",
  publisher = "American Physical Society (APS)",
  volume    =  7,
  number    =  7,
  month     =  "11~" # jul,
  year      =  2022,
  url       = "http://dx.doi.org/10.1103/physrevfluids.7.l071101",
  doi       = "10.1103/physrevfluids.7.l071101",
  issn      = "2469-990X,2469-9918"
}

@ARTICLE{Dehkharghani2023-mw,
  title     = "Self-transport of swimming bacteria is impaired by porous
               microstructure",
  author    = "Dehkharghani, Amin and Waisbord, Nicolas and Guasto, Jeffrey S",
  journal   = "Commun. Phys.",
  publisher = "Springer Science and Business Media LLC",
  volume    =  6,
  number    =  1,
  pages     = "1--9",
  month     =  "24~" # jan,
  year      =  2023,
  url       = "http://dx.doi.org/10.1038/s42005-023-01136-w",
  doi       = "10.1038/s42005-023-01136-w",
  issn      = "2399-3650,2399-3650"
}

@ARTICLE{Son2013-os,
  title    = "Bacteria can exploit a flagellar buckling instability to change
              direction",
  author   = "Son, Kwangmin and Guasto, Jeffrey S and Stocker, Roman",
  journal  = "Nat. Phys.",
  volume   =  9,
  number   =  8,
  pages    = "494--498",
  year     =  2013,
  url      = "http://dx.doi.org/10.1038/nphys2676",
  doi      = "10.1038/nphys2676",
  issn     = "1745-2473,1745-2481"
}

@ARTICLE{Guadayol2017-od,
  title    = "Cell morphology governs directional control in swimming bacteria",
  author   = "Guadayol, Òscar and Thornton, Katie L and Humphries, Stuart",
  journal  = "Sci. Rep.",
  volume   =  7,
  number   =  1,
  pages    =  2061,
  year     =  2017,
  url      = "http://dx.doi.org/10.1038/s41598-017-01565-y",
  doi      = "10.1038/s41598-017-01565-y",
  issn     = "2045-2322"
}

@ARTICLE{Bhattacharjee2021-oc,
  title    = "Chemotactic migration of bacteria in porous media",
  author   = "Bhattacharjee, Tapomoy and Amchin, Daniel B and Ott, Jenna A and
              Kratz, Felix and Datta, Sujit S",
  journal  = "Biophys. J.",
  volume   =  120,
  number   =  16,
  pages    = "3483--3497",
  year     =  2021,
  url      = "http://dx.doi.org/10.1016/j.bpj.2021.05.012",
  doi      = "10.1016/j.bpj.2021.05.012",
  issn     = "0006-3495,1542-0086"
}

@ARTICLE{De_Anna2021-io,
  title    = "Chemotaxis under flow disorder shapes microbial dispersion in
              porous media",
  author   = "de Anna, Pietro and Pahlavan, Amir A and Yawata, Yutaka and
              Stocker, Roman and Juanes, Ruben",
  journal  = "Nat. Phys.",
  volume   =  17,
  number   =  1,
  pages    = "68--73",
  year     =  2021,
  url      = "http://dx.doi.org/10.1038/s41567-020-1002-x",
  doi      = "10.1038/s41567-020-1002-x",
  issn     = "1745-2473,1745-2481"
}

@ARTICLE{Kaya2012-po,
  title    = "{Direct Upstream Motility in Escherichia coli}",
  author   = "Kaya, Tolga and Koser, Hur",
  journal  = "Biophys. J.",
  volume   =  102,
  number   =  7,
  pages    = "1514--1523",
  year     =  2012,
  url      = "http://dx.doi.org/10.1016/j.bpj.2012.03.001",
  doi      = "10.1016/j.bpj.2012.03.001",
  issn     = "0006-3495,1542-0086"
}

@ARTICLE{Drescher2013-xn,
  title    = "Biofilm streamers cause catastrophic disruption of flow with
              consequences for environmental and medical systems",
  author   = "Drescher, Knut and Shen, Yi and Bassler, Bonnie L and Stone,
              Howard A",
  journal  = "Proc. Natl. Acad. Sci. U. S. A.",
  volume   =  110,
  number   =  11,
  pages    = "4345--4350",
  month    =  "12~" # mar,
  year     =  2013,
  url      = "http://dx.doi.org/10.1073/pnas.1300321110",
  doi      = "10.1073/pnas.1300321110",
  pmc      = "PMC3600445",
  pmid     =  23401501,
  issn     = "0027-8424,1091-6490"
}

@ARTICLE{Torres_Maldonado2024-wu,
  title    = "Enhancement of bacterial rheotaxis in non-Newtonian fluids",
  author   = "Torres Maldonado, Bryan O and Théry, Albane and Tao, Ran and
              Brosseau, Quentin and Mathijssen, Arnold J T M and Arratia, Paulo
              E",
  journal  = "Proc. Natl. Acad. Sci. U. S. A.",
  volume   =  121,
  number   =  50,
  pages    = "e2417614121",
  year     =  2024,
  url      = "http://dx.doi.org/10.1073/pnas.2417614121",
  doi      = "10.1073/pnas.2417614121",
  issn     = "0027-8424,1091-6490"
}

@ARTICLE{Philip1970-rf,
  title    = "Flow in Porous Media",
  author   = "Philip, J R",
  journal  = "Annu. Rev. Fluid Mech.",
  volume   =  2,
  number   =  1,
  pages    = "177--204",
  year     =  1970,
  url      = "http://dx.doi.org/10.1146/annurev.fl.02.010170.001141",
  doi      = "10.1146/annurev.fl.02.010170.001141",
  issn     = "0066-4189,1545-4479"
}

@ARTICLE{Guasto2012-da,
  title    = "{Fluid Mechanics of Planktonic Microorganisms}",
  author   = "Guasto, Jeffrey S and Rusconi, Roberto and Stocker, Roman",
  journal  = "Annu. Rev. Fluid Mech.",
  volume   =  44,
  number   =  1,
  pages    = "373--400",
  year     =  2012,
  url      = "http://dx.doi.org/10.1146/annurev-fluid-120710-101156",
  doi      = "10.1146/annurev-fluid-120710-101156",
  issn     = "0066-4189,1545-4479"
}

@ARTICLE{Cao2024-nm,
  title    = "Giant enhancement of bacterial upstream swimming in macromolecular
              flows",
  author   = "Cao, Ding and Tao, Ran and Théry, Albane and Liu, Song and
              Mathijssen, Arnold J T M and Wu, Yilin",
  journal  = "Under Review",
  pages    = "arXiv:2408.13694",
  year     =  2024,
  url      = "http://arxiv.org/abs/2408.13694"
}

@ARTICLE{Baveye2017-bc,
  title    = "Microbial competition and evolution in natural porous
              environments: Not that simple",
  author   = "Baveye, Philippe C and Darnault, Christophe",
  journal  = "Proc. Natl. Acad. Sci. U. S. A.",
  volume   =  114,
  number   =  14,
  year     =  2017,
  url      = "http://dx.doi.org/10.1073/pnas.1700992114",
  doi      = "10.1073/pnas.1700992114",
  issn     = "0027-8424,1091-6490"
}

@ARTICLE{Adler1988-qy,
  title    = "Multiphase Flow in Porous Media",
  author   = "Adler, P M and Brenner, H",
  journal  = "Annu. Rev. Fluid Mech.",
  volume   =  20,
  number   =  1,
  pages    = "35--59",
  year     =  1988,
  url      = "http://dx.doi.org/10.1146/annurev.fl.20.010188.000343",
  doi      = "10.1146/annurev.fl.20.010188.000343",
  issn     = "0066-4189,1545-4479"
}

@ARTICLE{Conrad2012-xr,
  title    = "{Physics of bacterial near-surface motility using flagella and
              type IV pili: Implications for biofilm formation}",
  author   = "Conrad, Jacinta C",
  journal  = "Res. Microbiol.",
  volume   =  163,
  number   = "9-10",
  pages    = "619--629",
  year     =  2012,
  url      = "http://dx.doi.org/10.1016/J.RESMIC.2012.10.016",
  doi      = "10.1016/J.RESMIC.2012.10.016",
  issn     = "0923-2508,1769-7123"
}

@ARTICLE{Hallatschek2023-zr,
  title    = "Proliferating active matter",
  author   = "Hallatschek, Oskar and Datta, Sujit S and Drescher, Knut and
              Dunkel, Jörn and Elgeti, Jens and Waclaw, Bartek and Wingreen, Ned
              S",
  journal  = "Nat. Rev. Phys.",
  volume   =  5,
  number   =  7,
  pages    = "407--419",
  year     =  2023,
  url      = "http://dx.doi.org/10.1038/s42254-023-00593-0",
  doi      = "10.1038/s42254-023-00593-0",
  issn     = "2522-5820"
}

@ARTICLE{Di_Leonardo2011-cz,
  title    = "{Swimming with an image}",
  author   = "Di Leonardo, R and Dell'Arciprete, D and Angelani, L and Iebba, V
              and Dell'Arciprete, D and Angelani, L and Iebba, V and
              Dell'Arciprete, D and Angelani, L and Iebba, V",
  journal  = "Phys. Rev. Lett.",
  volume   =  106,
  number   =  3,
  pages    =  038101,
  year     =  2011,
  url      = "http://dx.doi.org/10.1103/PhysRevLett.106.038101",
  doi      = "10.1103/PhysRevLett.106.038101",
  issn     = "0031-9007,1079-7114"
}

@ARTICLE{Persat2015-rl,
  title    = "The Mechanical World of Bacteria",
  author   = "Persat, Alexandre and Nadell, Carey D and Kim, Minyoung Kevin and
              Ingremeau, Francois and Siryaporn, Albert and Drescher, Knut and
              Wingreen, Ned S and Bassler, Bonnie L and Gitai, Zemer and Stone,
              Howard A",
  journal  = "Cell",
  volume   =  161,
  number   =  5,
  pages    = "988--997",
  year     =  2015,
  url      = "http://dx.doi.org/10.1016/j.cell.2015.05.005",
  doi      = "10.1016/j.cell.2015.05.005",
  issn     = "0092-8674,1097-4172"
}

@ARTICLE{Dentz2022-yt,
  title     = "Dispersion of motile bacteria in a porous medium",
  author    = "Dentz, Marco and Creppy, Adama and Douarche, Carine and Clément,
               Eric and Auradou, Harold",
  journal   = "J. Fluid Mech.",
  publisher = "Cambridge University Press (CUP)",
  volume    =  946,
  number    = "A33",
  month     =  "10~" # sep,
  year      =  2022,
  url       = "http://dx.doi.org/10.1017/jfm.2022.596",
  doi       = "10.1017/jfm.2022.596",
  issn      = "0022-1120,1469-7645"
}

@ARTICLE{Jaimes-Lizcano2014-sn,
  title     = "Filamentous Escherichia coli cells swimming in tapered
               microcapillaries",
  author    = "Jaimes-Lizcano, Yuly A and Hunn, Dayton D and Papadopoulos,
               Kyriakos D",
  journal   = "Res. Microbiol.",
  publisher = "Elsevier BV",
  volume    =  165,
  number    =  3,
  pages     = "166--174",
  month     =  apr,
  year      =  2014,
  url       = "http://dx.doi.org/10.1016/j.resmic.2014.01.007",
  doi       = "10.1016/j.resmic.2014.01.007",
  pmid      =  24566556,
  issn      = "0923-2508,1769-7123"
}

@ARTICLE{Das2025-qb,
  title         = "Vorticity-induced surfing and trapping in porous media",
  author        = "Das, Pallabi and Residori, Mirko and Voigt, Axel and Mandal,
                   Suvendu and Kurzthaler, Christina",
  journal       = "arXiv [cond-mat.soft]",
  month         =  "4~" # nov,
  year          =  2025,
  url           = "http://dx.doi.org/10.48550/arXiv.2511.02471",
  archivePrefix = "arXiv",
  primaryClass  = "cond-mat.soft",
  eprint        = "2511.02471",
  doi           = "10.48550/arXiv.2511.02471"
}

@ARTICLE{Chang2018-ef,
  title     = "Surface topography hinders bacterial surface motility",
  author    = "Chang, Yow-Ren and Weeks, Eric R and Ducker, William A",
  journal   = "ACS Appl. Mater. Interfaces",
  publisher = "American Chemical Society (ACS)",
  volume    =  10,
  number    =  11,
  pages     = "9225--9234",
  month     =  "21~" # mar,
  year      =  2018,
  url       = "http://dx.doi.org/10.1021/acsami.7b16715",
  doi       = "10.1021/acsami.7b16715",
  issn      = "1944-8244,1944-8252"
}

@ARTICLE{Maki2000-kp,
  title     = "Motility and chemotaxis of filamentous cells of Escherichia coli",
  author    = "Maki, N and Gestwicki, J E and Lake, E M and Kiessling, L L and
               Adler, J",
  journal   = "J. Bacteriol.",
  publisher = "American Society for Microbiology",
  volume    =  182,
  number    =  15,
  pages     = "4337--4342",
  month     =  aug,
  year      =  2000,
  url       = "http://dx.doi.org/10.1128/JB.182.15.4337-4342.2000",
  doi       = "10.1128/JB.182.15.4337-4342.2000",
  pmc       = "PMC101954",
  pmid      =  10894745,
  issn      = "0021-9193,1098-5530"
}

@ARTICLE{Datta2025-uu,
  title     = "Bacterial swimming in porous gels exhibits intermittent run
               motility with active turns and mechanical trapping",
  author    = "Datta, Agniva and Beier, Sönke and Pfeifer, Veronika and
               Großmann, Robert and Beta, Carsten",
  journal   = "Sci. Rep.",
  publisher = "Springer Science and Business Media LLC",
  volume    =  15,
  number    =  1,
  pages     =  20320,
  month     =  "27~" # jun,
  year      =  2025,
  url       = "http://dx.doi.org/10.1038/s41598-025-02741-1",
  doi       = "10.1038/s41598-025-02741-1",
  pmc       = "PMC12205082",
  pmid      =  40579453,
  issn      = "2045-2322"
}

@ARTICLE{Perez-Estay2024-ze,
  title     = "Accumulation and depletion of {E}. coli in surfaces mediated by
               curvature",
  author    = "Pérez-Estay, Benjamín and Cordero, María Luisa and Sepúlveda,
               Néstor and Soto, Rodrigo",
  journal   = "Phys. Rev. E.",
  publisher = "American Physical Society (APS)",
  volume    =  109,
  number    = "5-1",
  pages     =  054601,
  month     =  may,
  year      =  2024,
  url       = "http://dx.doi.org/10.1103/PhysRevE.109.054601",
  doi       = "10.1103/PhysRevE.109.054601",
  pmid      =  38907493,
  issn      = "2470-0053,2470-0045"
}

@ARTICLE{Gonzalez-La-Corte2025-zi,
  title     = "Morphogenesis of bacterial cables in polymeric environments",
  author    = "Gonzalez La Corte, Sebastian and Stevens, Corey A and
               Cárcamo-Oyarce, Gerardo and Ribbeck, Katharina and Wingreen, Ned
               S and Datta, Sujit S",
  journal   = "Sci. Adv.",
  publisher = "American Association for the Advancement of Science (AAAS)",
  volume    =  11,
  number    =  3,
  pages     = "eadq7797",
  month     =  "17~" # jan,
  year      =  2025,
  url       = "http://dx.doi.org/10.1126/sciadv.adq7797",
  doi       = "10.1126/sciadv.adq7797",
  pmc       = "PMC11740958",
  pmid      =  39823332,
  issn      = "2375-2548"
}

@ARTICLE{Higashitani1995-qn,
  title    = "A cell division inhibitor {SulA} of Escherichia coli directly
              interacts with {FtsZ} through {GTP} hydrolysis",
  author   = "Higashitani, A and Higashitani, N and Horiuchi, K",
  journal  = "Biochem. Biophys. Res. Commun.",
  volume   =  209,
  number   =  1,
  pages    = "198--204",
  month    =  "6~" # apr,
  year     =  1995,
  url      = "http://dx.doi.org/10.1006/bbrc.1995.1489",
  doi      = "10.1006/bbrc.1995.1489",
  pmid     =  7726836,
  issn     = "0006-291X,1090-2104"
}

@ARTICLE{Esteves2025-jg,
  title     = "Nitric oxide promotes rapid development of motility to accelerate
               biofilm dispersal in Vibrio cholerae",
  author    = "Esteves, Nathaniel C and Tao, Ran and Pu, Qinqin and Banerjee,
               Arkaprabha and Mathijssen, Arnold J T M and Zhu, Jun",
  journal   = "Proc. Natl. Acad. Sci. U. S. A.",
  publisher = "Proceedings of the National Academy of Sciences",
  volume    =  122,
  number    =  49,
  pages     = "e2526864122",
  month     =  "9~" # dec,
  year      =  2025,
  url       = "http://dx.doi.org/10.1073/pnas.2526864122",
  doi       = "10.1073/pnas.2526864122",
  pmid      =  41329732,
  issn      = "0027-8424,1091-6490"
}

@ARTICLE{Zhu2025-as,
  title     = "Propulsion contribution from individual filament in a flagellar
               bundle",
  author    = "Zhu, Jin and Qiao, Yateng and Yan, Lingchun and Zeng, Yan and Wu,
               Yibo and Bian, Hongyi and Huang, Yidi and Ye, Yuxin and Huang,
               Yingyue and Hii, Russell Ching Wei and Teng, Yinuo and Guo,
               Yunlong and Li, Gaojin and Qu, Zijie",
  journal   = "Appl. Phys. Lett.",
  publisher = "AIP Publishing",
  volume    =  126,
  number    =  7,
  pages     =  073702,
  month     =  "17~" # feb,
  year      =  2025,
  url       = "http://dx.doi.org/10.1063/5.0243416",
  doi       = "10.1063/5.0243416",
  issn      = "0003-6951,1077-3118"
}

@ARTICLE{Aranson2006-rt,
  title     = "Patterns and collective behavior in granular media: Theoretical
               concepts",
  author    = "Aranson, Igor S and Tsimring, Lev S",
  journal   = "Rev. Mod. Phys.",
  publisher = "American Physical Society (APS)",
  volume    =  78,
  number    =  2,
  pages     = "641--692",
  month     =  "1~" # jun,
  year      =  2006,
  url       = "http://dx.doi.org/10.1103/revmodphys.78.641",
  doi       = "10.1103/revmodphys.78.641",
  issn      = "0034-6861,1539-0756"
}

@ARTICLE{Porter2025-ah,
  title     = "On the growth and form of bacterial colonies",
  author    = "Porter, Rachel and Trenado-Yuste, Carolina and Martinez-Calvo,
               Alejandro and Su, Morgan and Wingreen, Ned S and Datta, Sujit S
               and Huang, Kerwyn Casey",
  journal   = "Nat. Rev. Phys.",
  publisher = "Springer Science and Business Media LLC",
  volume    =  7,
  number    =  10,
  pages     = "535--553",
  month     =  "1~" # sep,
  year      =  2025,
  url       = "https://www.nature.com/articles/s42254-025-00849-x",
  doi       = "10.1038/s42254-025-00849-x",
  issn      = "2522-5820"
}

@ARTICLE{Tao2025-mi,
  title     = "Invasion of bacteria swimming upstream into microstructured
               devices",
  author    = "Tao, Ran and Théry, Albane and Que, Suya and Mathijssen, Arnold J
               T",
  journal   = "Newton",
  publisher = "Elsevier",
  volume    =  0,
  number    =  0,
  month     =  "18~" # dec,
  year      =  2025,
  url       = "http://dx.doi.org/10.1016/j.newton.2025.100337",
  doi       = "10.1016/j.newton.2025.100337",
  issn      = "2950-6360"
}


\newpage


\end{document}